\documentclass[graybox, envcountchap, authoryear, natbib]{svmult}

\usepackage{mathptmx}       
\usepackage{helvet}         
\usepackage{courier}        
\usepackage{type1cm}        
\usepackage{graphicx}        
\usepackage{multicol}        
\usepackage[bottom]{footmisc}
\usepackage{hyperref}
\hypersetup{colorlinks,citecolor=blue}

\newcommand{\apg}{\:^{>}_{\sim}\:}
\newcommand{\apl}{\:^{<}_{\sim}\:}

\newcommand{\msol}{\mbox{${\rm M}_\odot$}}
\newcommand{\mstar}{\mbox{$M_{\rm star}$}}

\newcommand{\kms}{\,${\rm{km\,s}^{-1}}$}
\newcommand{\cmjj}{\mbox{${\rm cm^{-2}}$}}
\newcommand{\HI}{\mbox{${\rm H\,{\scriptsize I}}$}}
\newcommand{\lya}{\mbox{${\rm Ly}\alpha$}}

\newcommand{\mnras}{MNRAS} 
\newcommand{\aj}{AJ}
\newcommand{\apj}{ApJ} 
\newcommand{\apjl}{ApJL}
\newcommand{\apjs}{ApJS} 
\newcommand{\aap}{A\&A}
\newcommand{\araa}{ARA\&A} 
\newcommand{\jcap}{JCAP}
\newcommand{\nat}{Nature}
\newcommand{\rmxaa}{Revista Mexicana de Astronomía y Astrofísica}
\newcommand{\apss}{Astrophysics and Space Science}
\newcommand{\MgII}{{\mbox{Mg\,{\scriptsize II}\ }}}
\newcommand{\FeII}{{\mbox{Fe\,{\scriptsize II}\ }}}
\newcommand{\OII}{{\mbox{[O\,{\scriptsize II}]}}}
\newcommand{\OVI}{{\mbox{[O\,{\scriptsize VI}]}}}

\begin{document}

\title*{The Circumgalactic Medium in Massive Halos}
\author{Hsiao-Wen Chen}
\institute{Hsiao-Wen Chen \at Department of Astronomy \& Astrophysics, and Kavli Institute for
Cosmological Physics, The University of Chicago, Chicago, IL 60637, USA,  \email{hchen@oddjob.uchicago.edu}}
%
%
\maketitle

\abstract{This chapter presents a review of the current state of
  knowledge on the cool ($T\sim 10^4$ K) halo gas content around
  massive galaxies at $z\approx 0.2-2$.  Over the last decade,
  significant progress has been made in characterizing the cool
  circumgalactic gas in massive halos of $M_h \approx
  10^{12-14}\,\msol$ at intermediate redshifts using absorption
  spectroscopy.  Systematic studies of halo gas around massive
  galaxies beyond the nearby universe are made possible by large
  spectroscopic samples of galaxies and quasars in public archives.
  In addition to accurate and precise constraints for the incidence of
  cool gas in massive halos, detailed characterizations of gas
  kinematics and chemical compositions around massive quiescent
  galaxies at $z\approx 0.5$ have also been obtained.  Combining all
  available measurements shows that infalling clouds from external
  sources are likely the primary source of cool gas detected at $d\apg
  100$ kpc from massive quiescent galaxies.  The origin of the gas
  closer in is currently less certain, but SNe~Ia driven winds appear
  to contribute significantly to cool gas found at $d< 100$ kpc.  In
  contrast, cool gas observed at $d\apl 200$ kpc from luminous quasars
  appears to be intimately connected to quasar activities on parsec
  scales.  The observed strong correlation between cool gas covering
  fraction in quasar host halos and quasar bolometric luminosity
  remains a puzzle.  Combining absorption-line studies with
  spatially-resolved emission measurements of both gas and galaxies is
  the necessary next step to address remaining questions.}

\section{Introduction}
\label{sec:intro}

In the standard picture of galaxy formation and evolution, primordial
gas first cools and condenses within dark matter halos to form stars
\citep[e.g.][]{White:1978, Blumenthal:1984}.  Gas in high-mass halos
with dark matter halo mass exceeding $M_h \approx 10^{12}\,\msol$ is
expected to be shock-heated to high temperatures
\citep[e.g.][]{Birnboim:2003, Keres:2005, Keres:2009, Dekel:2006}. Within the
hot gas halos, thermal instabilities can induce the formation of
pressure-supported cool clouds \citep[e.g.][]{Mo:1996, Maller:2004,
  Sharma:2012, Voit:2015}.  In lower mass halos, cool filaments from
the intergalactic medium can reach deep into the center of the halo
without being shock-heated.  Both condensed cool clouds of hot halos
and cool filaments can, in principle, supply the fuels necessary to
sustain star formation in galaxies.  As new stars form and evolve, the
surrounding interstellar and circumstellar gas is expected to be
heated and enriched by heavy elements ejected from massive stars,
regulating subsequent star formation.

Several semi-analytic studies have been carried out in searching for a
general prescription that connects galaxies found in observations to
dark matter halos formed in theoretical frameworks.  These studies
seek to establish a mean relation between galaxy stellar mass ($M_{\rm
  star}$) and host halo mass ($M_h$) by matching the observed space
density of galaxies with the expected abundance of dark matter halos
as a function of mass \citep[e.g.][]{Vale:2004,Shankar:2006}.  The mean
galaxy mass and halo mass ratio is found to peak at $M_h\sim10^{12}\,\msol$ with $M_{\rm star}/M_h\approx 0.04$ and declines
rapidly both toward higher and lower masses
\citep[e.g.][]{Moster:2010,Guo:2010,Behroozi:2010}.  The declining $M_{\rm
  star}/M_h$ indicates a reduced star formation efficiency in both
low- and high-mass halos.  Different feedback mechanisms are invoked
in theoretical models in order to match the observed low star
formation efficiency in low- and high-mass halos.  As supernova-driven
winds are thought to suppress star formation efficiency in low-mass
dwarf galaxies \citep[e.g.][]{Larson:1974, Dekel:1986}, feedback due to
active galactic nuclei (AGN) powered by supermassive black holes is
invoked to quench star formation in high-mass halos, resulting in
massive quiescent galaxies 
\citep[e.g.][]{Bower:2006,Croton:2006,Dubois:2013}.  While blueshifted broad
absorption and emission lines are commonly seen in luminous quasars,
indicating the presence of high-speed outflows, direct observational
evidence of AGN feedback on large scales ($\sim 10-100$ kpc) remains
scarce \citep[e.g.][]{Alexander:2010,Greene:2012,Maiolino:2012} and see
also \cite{Fabian:2012} for a review.

Observations of the circumgalactic medium (CGM) in massive halos offer
complementary and critical constraints for the extent of feedback and
gas accretion \citep[e.g.][]{Somerville:2015}.  In particular, the
circumgalactic space within the halo radius, $R_{\rm vir}$, lies
between galaxies, where star formation takes place, and the
intergalactic medium (IGM), where 90 percent of all baryonic matter in
the universe resides \citep{Rauch:1997}.  As a result, CGM properties
are shaped by the complex interactions between IGM accretion and
outflows driven by energetic feedback processes in the galaxies.

Imaging observations of the cool ($T\sim 10^4$ K) CGM are only
feasible in 21~cm surveys at $z\apl 0.2$, because, with few exceptions
\citep[e.g.][]{Cantalupo:2014, Fernandez:2016, Hayes:2016}, the gas
density is typically too low to be detected in emission.
Absorption-line spectroscopy of background quasars provides a
powerful, alternative tool for studying this tenuous gas in the
distant universe based on the absorption features imprinted in the
quasar spectra 
\citep[e.g.][]{Lanzetta:1995,Bowen:1995,Steidel:2002}.
But, because both quasars and massive galaxies are rare, close pairs
of massive galaxies and background quasars by chance projection are
even rarer.  Studying the CGM around massive galaxies using quasar
absorption spectroscopy therefore requires a large spectroscopic
sample of galaxies and quasars over a substantial volume in order to
assemble a statistical sample of massive galaxy and quasar pairs.  The
Sloan Digital Sky Survey \citep[SDSS;][]{York:2000} has produced a large
spectroscopic archive of distant galaxies and quasars, facilitating
the assembly of a statistically significant sample of close massive
galaxy and quasar pairs, as well as a large sample of projected quasar
pairs.  These pair samples have enabled systematic studies of halo gas
beyond the nearby universe using absorption spectroscopy.

This chapter presents a review of the current state of knowledge on
the cool CGM properties in massive halos of $M_h \approx
10^{12-14}\,\msol$ at $z\approx 0.2-2$.  Specifically, the review will
focus on massive quiescent galaxies with $\mstar\apg 10^{11}\,\msol$
at $z\apl 1$, with additional coverage on quasar host halos.
Empirical studies of cool/warm gas in galaxy groups and clusters of
$M_h \apg 10^{14}\,\msol$ have been carried out for a small sample
\citep[e.g.][]{Lopez:2008, Yoon:2012, Andrews:2013, Stocke:2014}, but a
detailed understanding of the gas phase in galaxy cluster and group
environments relies primarily on x-ray studies of the hot plasma.
Extensive reviews on the x-ray properties of intragroup and
intracluster gas can be found in \cite{Mulchaey:2000, Mathews:2003,
  Kravtsov:2012}.

The emphasis on quiescent galaxies in high-mass halos is motivated by
the empirical finding that more than 90\% of massive galaxies with
$\mstar\apg 10^{11}\,\msol$ in the local universe contain primarily
evolved stellar populations with little on-going star formation
\citep[e.g.][]{Peng:2010,Tinker:2013}.  It is therefore expected that a general
understanding of the CGM properties in massive halos can be
established based on observations of massive quiescent galaxies.  An
added bonus in studying massive quiescent galaxies is the unique
opportunity to explore other feedback mechanisms for quenching star
formation in massive halos, in the absence of complicated starburst
driven winds.

The emphasis on quasar host halos is motivated by two factors.  First,
while determining quasar host mass is difficult due to uncertain host
galaxy properties, the large observed clustering amplitude indicates
that the mean halo mass of quasar hosts is high, $M_h\apg
10^{12.5}\,\msol$ \citep{Porciani:2004, White:2012, Shen:2013Q}.  In
addition, observations of the CGM properties in quasar host halos
directly address the issues concerning the extent of AGN feedback
\citep[e.g.][]{Shen:2013, Fumagalli:2014, Rahmati:2015, Faucher:2015,
  Faucher:2016}.  Most nearby elliptical galaxies are found to host a
supermassive black hole at the center \citep{Kormendy:1995,Ho:2008},
and therefore high-redshift quasar hosts are likely the progenitors of
these nearby massive quiescent galaxies.  Studies of the CGM in quasar
host halos may provide important clues for how the CGM properties are
shaped while the galaxy undergoes an active quasar phase.

\section{Incidence/Covering Fraction of Cool Gas in Quiescent Halos}
\label{sec:lrg}

\begin{figure}[t]
\begin{center}
\includegraphics[width=4.5in]{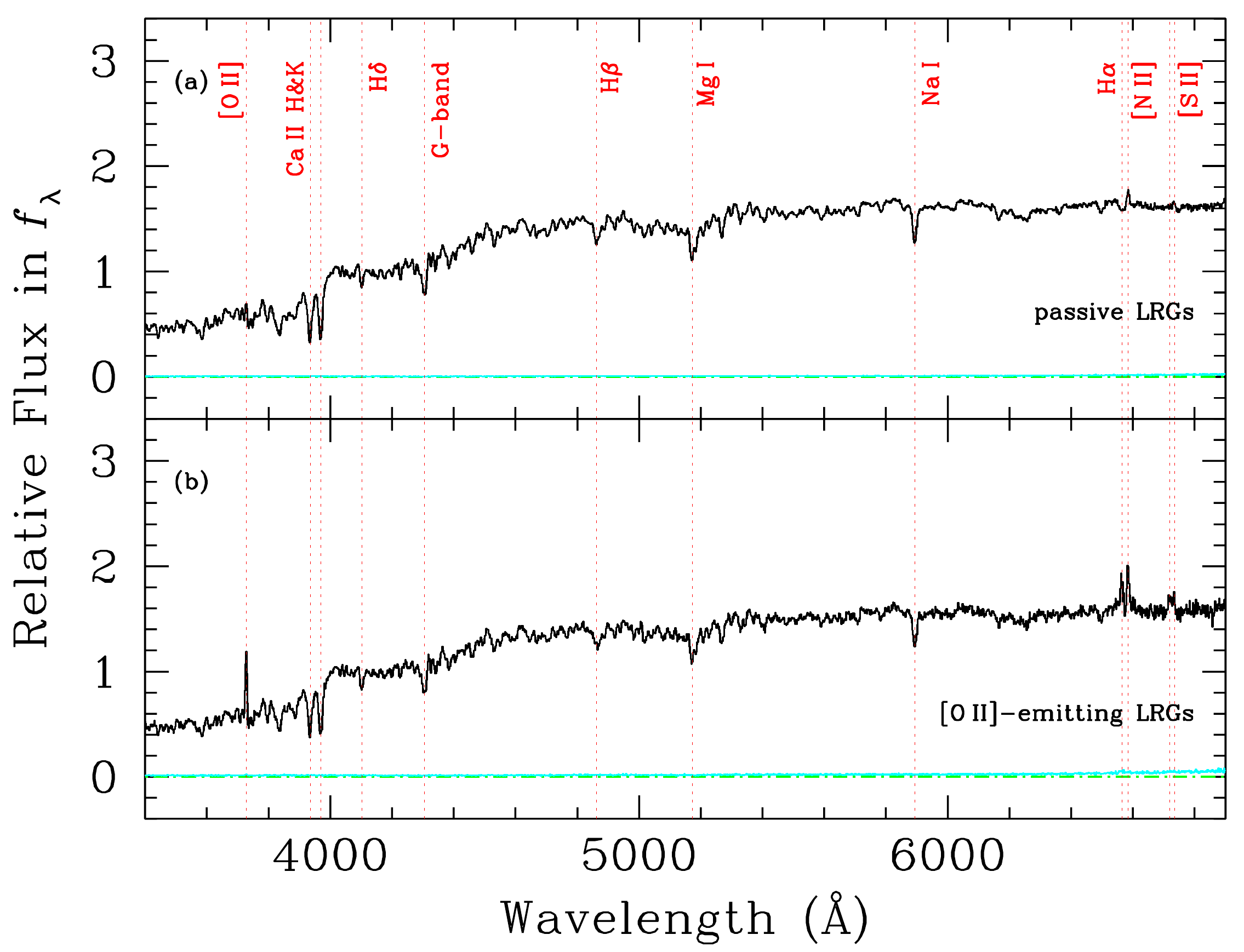}
\end{center}
\caption{The quiescent nature of $z\approx 0.5$ luminous red galaxies
  (LRGs) shown in stacked rest-frame optical spectra.  Panel (a) shows
  the median stack of $\sim 1780$ passive LRGs and panel (b) shows the
  median stack of $\sim 280$ LRGs with [O\,II] emission detected (see
  \cite{Huang:2016} for a detailed description of the sample
  definition and selection).  Prominent absorption and emission lines
  are labeled in red.  The corresponding 1-$\sigma$ dispersion in each
  stack is shown in cyan at the bottom of each panel.  In both stacked
  spectra, the spectral properties are characterized by prominent
  absorption features due to Ca\,II H and K, G-band, Mg\,I and Na\,I
  that indicate a predominantly old stellar population, as well as a
  relatively weak Balmer absorption series.  The observed high
  [N\,II]/H$\alpha$ emission ratios seen in both passive and
  [O\,II]-emitting LRGs, together with the low [O\,III]/[O\,II]
  ratios, suggest the presence of LINER in these LRGs, which could be
  attributed to the presence of underlying AGNs or evolved AGB stars
  in these galaxies.}
\label{fig:LRGspec}       
\end{figure}

Luminous red galaxies (LRGs) uncovered in the SDSS exhibit little
on-going star formation and display photometric and spectral
properties resembling nearby elliptical galaxies
\citep{Eisenstein:2001}.  They are characterized by a mean luminosity
of $\approx 5\,L_*$ and a mean stellar mass of 
$M_{\rm star}\approx 3 \times 10^{11}\,\msol$ at  
$z\approx 0.5$ \citep[e.g.][]{Tojeiro:2011},
with corresponding dark matter halo mass of 
$M_h \approx 3\times10^{13}\,\msol$ 
\citep[e.g.][]{Zheng:2007, Blake:2008, Padmanabhan:2009}, 
and therefore offer an ideal laboratory for
studying the cool gas content in massive quiescent halos.

Roughly 10\% of LRGs exhibit [O\,II] emission features
\citep{Eisenstein:2003, Roseboom:2006}, suggesting the possible
presence of on-going star formation in some of these massive galaxies.
However, a more detailed examination of their spectral properties
reveals unusually high [N\,II]/H$\alpha$ together with low
[O\,III]/[O\,II] ratios, both of which indicate the presence of
low-ionization nuclear emission line regions (LINER) in these galaxies
\citep[e.g.][]{Huang:2016}.  Figure \ref{fig:LRGspec} shows stacked
spectra of SDSS LRGs grouped into ``passive'' and [O\,II]-emitting
subsamples.  In both panels of Figure \ref{fig:LRGspec}, the stacked
spectra are characterized by prominent absorption features due to
Ca\,II H and K, G-band, Mg\,I and Na\,I, along with LINER-like
emission, namely strong [N\,II]/H$\alpha$ ratios.  The prominent
absorption features indicate a predominantly old stellar population,
whereas the LINER-like emission indicates the presence of additional
ionizing sources in these galaxies.  Although the origin of LINER-like
emission is not fully understood, the observed slow decline in the
spatial profile of LINER-like emission requires spatially distributed
ionizing sources, rather than centrally concentrated active galactic
nuclei (AGN), in these quiescent galaxies \citep[e.g.][]{Sarzi:2006,
  Sarzi:2010, Yan:2012, Singh:2013, Belfiore:2016}.  Likely sources
include photo-ionization due to hot, post-asymptotic giant branch
(post-AGB) stars or winds from Type Ia supernovae
\citep[e.g.][]{Conroy:2015}.  In the absence of on-going star formation, these
LRGs also provide a unique sample for testing additional feedback
mechanisms for quenching star formation in massive halos, including
gravitational heating \citep[e.g.][]{Johansson:2009} and winds from
evolved stars and Type Ia supernovae \citep[e.g.][]{Conroy:2015}.

A first step toward understanding the quiescent state of LRGs is to
characterize the incidence and covering fraction of cool gas in their
host halos.  Over the past several years, absorption-line observations
of background QSOs have revealed extended cool gas at projected
distances $d\approx 10-300$ kpc from LRGs at $z\sim 0.5$.
Specifically, the LRG--\MgII\ cross-correlation function displays an
amplitude that is comparable to the LRG auto-correlation on small
scales ($\apl 450$ comoving kpc), well within the virial radii of the
halos.  The large clustering amplitude on small scales suggests the
presence of \MgII\ gas inside these massive quiescent halos
\citep[e.g.][]{Bouche:2006, Tinker:2008, Gauthier:2009, Lundgren:2009}.  In
addition, surveys of \MgII\ $\lambda\lambda$\,2796, 2803 absorption
doublets in LRG halos have covered a non-negligible fraction of LRG
halos with associated \MgII\ absorbers \citep[e.g.][]{Bowen:2011,
  Gauthier:2011, Gauthier:2014, Zhu:2014, Huang:2016}.

These \MgII\ absorption transitions are commonly seen at projected
distances of $d\apl 100$\,kpc from star-forming galaxies
\citep[e.g.][]{Lanzetta:1990, Steidel:1994, Nestor:2007, Kacprzak:2008,
  Chen:2008, Chen:2010, Werk:2013} and are understood to arise
primarily in photo-ionized gas of temperature $T\sim 10^4$ K
\citep[e.g.][]{Bergeron:1986, Hamann:1997} and neutral hydrogen column density
$N(\HI)\approx 10^{18}-10^{22}$ \cmjj\ \citep[e.g.][]{Churchill:2000,
  Rao:2006}. The large rest wavelengths make the \MgII\ doublet a
convenient probe of chemically enriched gas at $z\approx 0.3-2.3$ in
optical QSO spectra.  In addition, the rest-frame absorption
equivalent width $W_r(2796)$ is found to increase proportionally with
the number of individual components and the velocity spread of the
components \citep[e.g.][]{Petitjean:1990, Churchill:2000,
  Churchill:2003}.  Measurements of $W_r(2796)$ therefore trace the
underlying gas kinematics corresponding to $\sim 100$ \kms\ per
Angstrom.  Several authors have proposed that strong \MgII\ absorbers
of $W_r(2796)>1$ \AA\ originate in starburst driven outflows
\citep[e.g.][]{Zibetti:2007, Menard:2011} and that the large $W_r(2796)$ is
driven by non-gravitational motion in the outflowing media 
\citep[e.g.][]{Bouche:2006}. The presence of strong \MgII\ absorbers near
quiescent galaxies is therefore particularly surprising.

\begin{figure}[!ht]
\begin{center}
\includegraphics[width=3.5in]{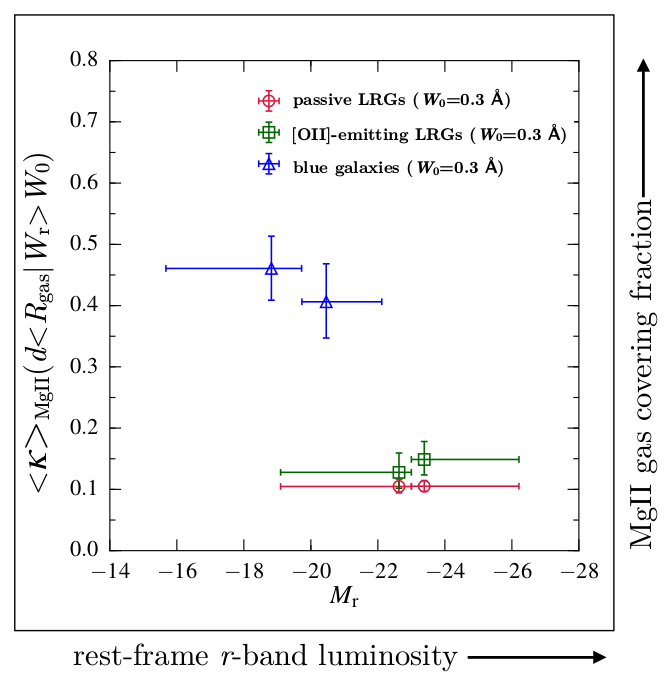}
\end{center}
%
%
\caption{Mass dependence of mean covering fraction of
  chemically-enriched cool gas in galaxy halos \citep{Chen:2010,
    Gauthier:2011, Huang:2016}.  The mean covering fraction,
  $\langle\kappa\rangle$, is calculated based on observations of
  \MgII\ absorbing gas within a fiducial halo gas radius, $R_{\rm
    gas}$.  Rest-frame $r$-band magnitude is adopted here as a proxy
  for the underlying total stellar mass \citep[see][]{Liang:2014}.
  Star-forming galaxies at $z\approx 0.3$ are shown in blue triangles,
  while luminous red galaxies (LRGs) at $z\approx 0.5$ are shown in
  squares/circles.  The covering fraction is determined for gas with
  absorption strength exceeding $W_0=0.3$ \AA, a detection threshold
  that is allowed by the quality of the absorption spectra.  The
  horizontal bars represent the full range of $M_r$ for galaxies
  included in each bin, and the vertical error bars represent the 68\%
  confidence interval.  In comparison to star-forming galaxies
  \citep{Chen:2010}, LRGs, being old and massive, exhibit a much
  reduced covering fraction of chemically-enriched cool gas as probed
  by the \MgII\ absorption features \citep{Gauthier:2011,Huang:2016}.
  However, observations also show a definitive presence of cool gas in
  these LRG halos, demonstrating that they are not completely devoid
  of star formation fuels \citep{Huang:2016}.  Furthermore, roughly
  10\% of LRGs exhibit [O\,II] emission features and LINER-like
  spectra.  These [O\,II]-emitting LRGs display a slightly elevated
  cool gas covering fraction from passive LRGs but the mean value
  remains significantly lower than what is seen in lower-mass,
  star-forming galaxies \citep{Huang:2016}.  Similar trends in the
  mean gas covering fraction have also been reported for H\,I
  \lya\ and high-ionization species probed by the
  O\,VI\,$\lambda\lambda$\,1031,1037 doublet transitions.  Massive
  quiescent galaxies also consistently display non-zero covering
  fractions of \lya\ and \OVI\ absorbing gas \citep{Tumlinson:2011,
    Thom:2012, Tumlinson:2013, Werk:2013}.}
\label{fig:Mg2cover}       
\end{figure}

Figure \ref{fig:Mg2cover} compares the observed mean covering fraction
of \MgII\ absorbing gas, $\langle\kappa\rangle_\MgII$, in halos around
typical $L_*$- and sub-$L_*$-type, star-forming galaxies (blue
triangles) with what is seen in LRG halos (green and red symbols) at
$z\approx 0.3-0.5$.  For a representative comparison across a broad
luminosity (or mass) range, the gas covering fraction is determined
within a fiducial halo gas radius, $R_{\rm gas}$, that scales with
galaxy luminosity according to $R_{\rm gas}\approx
107\,(L_B\,/\,L_{B}^*)^{0.35}$ kpc \citep{Kacprzak:2008, Chen:2010}.
In addition, $\kappa$ is estimated for halo clouds with absorption
strength exceeding $W_0=0.3$ \AA, a detection threshold that is
allowed by the quality of the absorption spectra.  Previous studies
have shown that typical $L_*$ galaxies have $R_{\rm gas}\approx 130$
kpc and $R_{\rm gas}$ scales with luminosity according to 
$R_{\rm gas}\propto L^{0.35}$ \citep[see][]{Chen:2008, Chen:2010}.  Following
this scaling relation, LRGs with a mean luminosity of $\approx
3.6\,L_*$ are expected to have $R_{\rm gas}\approx 206$ kpc.

Interpreting the rest-frame $r$-band absolute magnitude as a proxy for
the underlying stellar mass \citep[e.g.][]{Liang:2014}, Figure
\ref{fig:Mg2cover} displays a strong mass dependence in the observed
incidence and covering fraction of \MgII\ absorbing gas.
Specifically, the observed \MgII\ covering fraction in LRG halos is
more than four times lower than what is found in halos around
lower-mass, star-forming galaxies.  However, in spite of this
substantially reduced cool gas content in massive quiescent halos, the
distinctly non-zero \MgII\ gas covering fraction around LRGs
demonstrates the definitive presence of chemically-enriched cool gas
around evolved galaxies.  Finally, it is interesting to note that
while [O\,II]-emitting LRGs display a slightly elevated overall cool
gas content than passive LRGs, the mean gas covering fraction remains
under 20\% \citep{Huang:2016}.

Figure \ref{fig:Mg2cover} is based on two separate studies designed to
characterize the CGM properties of $L_*$/sub-$L_*$ galaxies using
$\sim 260$ galaxies at $z\approx 0.25$ \citep{Chen:2010} and those of
massive quiescent galaxies using $\sim 38000$ LRGs
\citep{Huang:2016}, respectively.  The COS-Halos survey was designed
for a systematic study of gaseous halos over a broad range of mass and
star formation history, and the sample contains 28 blue sub-$L_*$
galaxies with a median mass of $\langle\,M_{\rm star}({\rm
  blue})\,\rangle\approx 2\times 10^{10}\,\msol$ and 16 massive red
galaxies with a median mass of $\langle\,M_{\rm star}({\rm
  red})\,\rangle\approx 10^{11}\,\msol$ \citep{Tumlinson:2013,
  Werk:2013}.  While the uncertainties are large due to a smaller
sample size, the COS-Halos galaxies confirm the observed mass
dependence of \MgII\ gas covering fraction in Figure
\ref{fig:Mg2cover} \citep{Werk:2013}.

A declining gas covering fraction from low-mass, star-forming galaxies
to massive quiescent galaxies is also observed in H\,I and highly
ionized species probed by the O\,VI\,$\lambda\lambda$\,1031,1037
doublet transitions.  Specifically, the covering fraction of
moderately strong \lya\ absorbers of $W_r(1215)>0.1$ \AA\ declines
from $\apg 90$\% around blue, star-forming galaxies
\citep{Tumlinson:2013} to $\approx 40-50$\% around red passive
galaxies \citep{Thom:2012}, and the covering fraction of O\,VI
absorbing gas decreases from $\approx 90$\% to $\approx 30$\%
\citep{Tumlinson:2011, Werk:2013}.  In both cases, massive quiescent
galaxies consistently display non-zero covering fractions of \lya\ and
O\,VI absorbing gas.

Deep 21~cm and CO imaging surveys have also revealed that roughly 40\%
of elliptical galaxies in the nearby universe contain cool neutral gas
\citep[e.g.][]{Oosterloo:2010, Young:2014}.  While on-going star
formation is observed at low levels, $\apl 1\,\msol\,{\rm yr}^{-1}$, in
many early-type galaxies of $M_{\rm star}\approx 5\times
10^{10}\,\msol$, it is rarely seen in more massive systems of $M_{\rm
  star}\apg 10^{11}\,\msol$ \citep[e.g.][]{Salim:2010}. Combining
H\,I/CO imaging surveys of local elliptical galaxies and
absorption-line observations of distant early-type galaxies at
$z\approx 0.2-0.5$ demonstrates that cool gas is indeed present in
some, although not all, massive quiescent halos.

A declining cool gas fraction with increasing halo mass is expected in
theoretical models that attribute the observed \MgII\ absorbers to
infalling gas from either thermally unstable hot halos or the
intergalactic medium \citep[e.g.][]{Maller:2004, Keres:2009}.
In the presence of cool gas, it is expected that these halo clouds
could provide the fuels necessary for sustaining star formation in the
LRGs.  The lack of on-going star formation in these massive galaxies
over such an extended cosmological time period 
\citep[$\apg 2$ Gyr;][]{Gauthier:2011} therefore suggests a prolonged 
duty cycle for the underlying heat sources.


\section{Radial Profiles of Absorbing Gas}
\label{sec:spatial}

A key advantage of 21~cm and CO imaging surveys is their ability to
resolve the spatial distribution of the gas around individual
galaxies.  The morphologies of the detected H\,I and CO gas span a
broad range, from regular disk- or ring-like structures to irregular
distributions of clumps and/or streams \citep[e.g.][]{Oosterloo:2007,
  Serra:2012} with roughly 1/4 displaying centralized disk or
ring-like structures \citep{Serra:2012}.  Such a wide range of
morphologies suggests different origins of the gas in different
galaxies, including leftover materials from previous mergers and newly
accreted gas from the CGM/IGM.  On the other hand, QSO absorption
spectroscopy, while offering extremely sensitive probes of tenuous gas
beyond the limits of 21~cm/CO surveys, reveals gas properties along a
single sightline per halo and does not provide constraints for spatial
variations in individual halos.  Nevertheless, considering an ensemble
of close QSO and galaxy pairs over a wide range of projected
separations provides a spherical average of halo absorption properties
for the entire galaxy sample.

\begin{figure}[th]
\begin{center}
\includegraphics[width=4.5in]{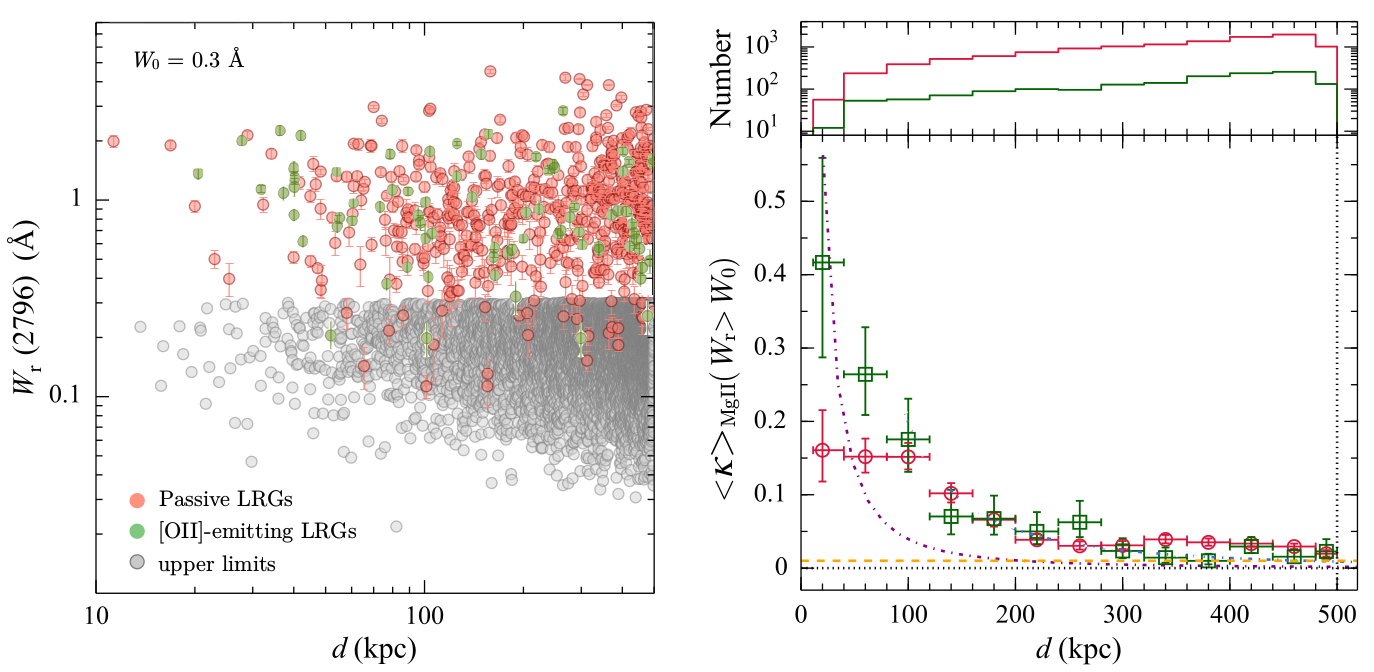}
\end{center}
%
%
\caption{Spatial distribution of the observed \MgII\ absorption
  strength ($W_r(2796)$; {\it left} panel) and covering fraction
  ($\langle\kappa\rangle_{\rm MgII}$; {\it right} panel) as a function
  of projected distance $d$ from LRGs at $z\approx 0.5$.  The figures
  are adapted from \cite{Huang:2016}.  The LRGs are
  selected to have a background QSO at $d<500$ kpc (the virial radius
  of a typical LRG halo) with sufficiently high signal-to-noise
  ($S/N$) absorption spectra available for constraining the
  presence/absence of \MgII\ absorbers with absorption strength
  exceeding $W_0=0.3$ \AA.  A total of 13330 LRGs in the SDSS data
  archive satisfy this criterion, 1575 of these exhibit
  [O\,II]-emission features and the rest are labeled as passive
  galaxies.  For non-detections, a 2-$\sigma$ upper limit of
  $W_r(2796)$ is presented for the LRG.  For a few sightlines, the QSO
  spectra are of sufficiently high $S/N$ for detecting \MgII\
  absorption lines as weak as $W_r(2796)\approx 0.1$ \AA\ (red and
  green data points with $W_r(2796)<0.3$ \AA\ in the left panel).
  These are by definition included in the covering fraction
  calculations in the right panel, although LRGs with $W_r(2796)<0.3$
  \AA\ are considered non-absorbing galaxies at the
  $W_0=0.3$-\AA\ threshold.  The upper-right panel displays the
  numbers of LRGs, passive (in red) and [O\,II]-emitting (in green),
  contributing to the $\langle\kappa\rangle_{\rm MgII}$ calculation in
  each 40-kpc bin (horizontal bars in the right panel).  Uncertainties
  in $\langle\kappa\rangle_{\rm MgII}$ are shown in vertical error
  bars, which represent the 68\% confidence intervals.  Possible
  contributions due to random background absorbers and gas-rich
  satellite galaxies are also shown in the orange dashed line and the
  purple dash-dotted curve, respectively.}
\label{fig:spatial}       
\end{figure}

Figure \ref{fig:spatial} shows the spatial distribution of the
observed \MgII\ absorption strength [$W_r(2796)$] and annular average
of \MgII\ gas covering fraction ($\langle\kappa\rangle_{\rm MgII}$) as
a function of projected distance ($d$) for a sample of 13330 LRGs at
$z\approx 0.5$ from \cite{Huang:2016}.  These LRGs are
selected to have a background QSO within $d\approx 500$ kpc, the
virial radius of a typical LRG halo, with sufficiently high
signal-to-noise ($S/N$) absorption spectra available in the SDSS
archive for detecting \MgII\ absorbers of $W_r(2796)\apg 0.3$ \AA.
The LRGs are grouped into passive and [O\,II]-emitting subsamples
based on the observed strength of [O\,II] emission lines (Figure
\ref{fig:LRGspec}).  A number of unique features are apparent in the
observed spatial distribution of chemically-enriched cool gas in LRG
halos:

\begin{itemize}

\item a relatively flat trend is seen in the observed $W_r(2796)$
  versus $d$ distribution for individual absorbers uncovered over the
  entire projected distance range $d\apl 500$ kpc of both passive and
  [O\,II]-emitting LRG halos, although the scatter is large ({\it
    left} panel);

\item a constant gas covering fraction,
  $\langle\kappa\rangle_{\MgII}\approx 15$\%, is found at $d<120$ kpc
  around passive LRGs, while $\langle\kappa\rangle_{\MgII}$ increases
  from $\langle\kappa\rangle_{\MgII}\approx 15$\% at $d\approx 120$
  kpc to $\langle\kappa\rangle_{\MgII}\approx 40$\% (though with a
  large error bar) at $d\apl 40$ kpc from [O\,II]-emitting LRGs;

\item the gas covering fraction declines rapidly at $d>120$ kpc,
  reaching the expected rate of incidence due to random background
  absorbers at $d\apg 300$ kpc (orange dashed line in the {\it right}
  panel);

\item strong \MgII\ absorbers of $W_r(2796)>1$ \AA\ appear to be
  common throughout the LRG halos at $d\apl 500$ kpc.

\end{itemize}

The observed flat $W_r(2796)$ versus $d$ trend for \MgII\ absorbers
and the large fraction of strong \MgII\ absorbers with $W_r(2796)>1$
\AA\ in LRG halos \citep[see also][]{Bowen:2011} are both in stark
contrast to what is known for star-forming, $L_*$ and sub-$L_*$
galaxies.  Specifically, CGM observations of star-forming galaxies at
both low and high redshifts have shown steadily declining absorption
strength, both in hydrogen and in heavy ions, with increasing
projected distance \citep[e.g.][]{Lanzetta:1990, Bowen:1995, Chen:1998,
  Chen:2001CIV, Steidel:2010, Liang:2014}.  Strong \MgII\ absorbers
with $W_r(2796)>1$ \AA\ are only found at $d \apl 60$ kpc from
isolated $L_*$ galaxies \citep[e.g.][]{Chen:2008, Chen:2010}.  In
addition, the extent of halo gas at a fixed absorption equivalent
width threshold, $R_{\rm gas}$, is found to scale with galaxy
luminosity \citep[e.g.][]{Chen:1998, Chen:2001HI, Kacprzak:2008,
  Chen:2008, Chen:2010}.  Therefore, the relation between $W_r(2796)$
versus $d/R_{\rm gas}$ exhibits a significantly smaller scatter than
the relation between $W_r(2796)$ and $d$.  This luminosity--$R_{\rm
  gas}$ scaling relation is understood as such that more luminous (and
presumably more massive) halos possess more extended halo gas.  For
the LRGs, however, the scatter in the observed $W_r(2796)$ versus
luminosity-normalized $d$ relation is the same as in the observed
$W_r(2796)$ versus $d$ distribution.  Figure \ref{fig:spatial} shows
that, aside from a declining $\langle\kappa_{\MgII}\rangle$ at $d>120$
kpc, the cool clouds that give rise to the observed \MgII\ absorption
features do not appear to depend on either the luminosity (or
equivalently mass) or projected distance of the LRGs\footnote{Using
  stacked QSO spectra at the rest frames of all LRGs, it has been
  shown that the mean absorption equivalent width
  $\langle\,W_0\,\rangle$ of \MgII\ gas declines steadily with
  increasing projected distance in LRG halos \citep{Zhu:2014}.  The
  steady decline of $\langle\,W_0\,\rangle$ does not necessarily
  contradict the relatively flat $W_r(2796)$ versus $d$ distribution
  for individual \MgII\ absorbers in Figure \ref{fig:spatial}, because
  $\langle\,W_0\,\rangle$ represents a weighted average of absorber
  strength and gas covering fraction.  A direct comparison between
  $\langle\,W_0\,\rangle$ and Figure \ref{fig:spatial} requires
  knowledge of the frequency distribution function of \MgII\ absorbers
  in LRG halos, $f(W_r)$, and the underlying number density ratio
  between absorbing and non-absorbing LRGs. }.

The lack of dependence of the observed \MgII absorbing clouds on the
LRG properties appears to rule out the possibility of these absorbers
being connected to outgoing materials from the LRGs.  However, the
observed flat mean covering fraction of \MgII absorbers at $d\apl 120$
kpc for passive LRGs is also inconsistent with the expectation for
accreted materials from the IGM, which are expected to show an
increasing covering fraction with decreasing projected distance
\citep[e.g.][]{Faucher:2011, Fumagalli:2011, Shen:2013, Ford:2014}.  
Such a discrepancy indicates that IGM accretion alone cannot explain the
spatial distribution of chemically-enriched cool gas in these massive
quiescent halos, at least not in the inner 120 kpc.

Contributions from the interstellar and circumgalactic gas of
satellite galaxies to the observed \MgII\ absorbers in LRG halos have
also been considered \citep[e.g.][]{Gauthier:2010, Huang:2016}.  In
particular, observations have shown that LRGs reside in overdense
environments with neighboring satellites \citep[e.g.][]{Tal:2012}.
Roughly 20\% of satellite galaxies are blue and presumably gas-rich
\citep[e.g.][]{Hansen:2009, Prescott:2011}.  Under the assumption that
these blue satellites can retain their gaseous halos, the expected
contributions to the observed covering fraction of \MgII\ gas can be
estimated based on the known luminosity scaling relation 
\citep[e.g.][]{Chen:2008}.  The expectation is shown in the purple curve in
the right panel of Figure \ref{fig:spatial}.  It is clear that blue
satellites alone cannot account for the observed 15\% covering
fraction of \MgII\-absorbing gas at $d \sim 100$ kpc in LRG halos.

A promising explanation for the observed \MgII\ absorbers in LRG halos
are cool clumps condensing out of thermally unstable hot halos 
\citep[e.g.][]{Mo:1996, Maller:2004, Voit:2015}.  In particular, hot halos are
a common feature among the LRG population, because efforts in search
of a Sunyaev-Zel'dovich signal \citep{Sunyaev:1970} around high-mass
LRGs ($M_h \apg 8\times 10^{13}\,\msol$) have yielded detections in
all mass bins studied \citep{Hand:2011}.  In addition, \MgII\
absorbers are frequently resolved into multiple components with the
number of components being proportional to both the rest-frame
absorption equivalent width $W_r(2796)$ \citep[e.g.][]{Petitjean:1990,
  Churchill:2003} and the minimum line-of-sight velocity spread
\citep[e.g.][]{Churchill:2000}.  The \MgII\ absorption equivalent width
is therefore driven by the line-of-sight cloud motion rather than by
the underlying total gas column density.  Recall also that the \MgII\
absorbers found in the vicinities of passive and [O\,II]-emitting LRGs
display a constant $W_r(2796)$ across the entire projected distance
range shown in Figure \ref{fig:spatial} with a mean and dispersion of
$\langle\,\log\,W_r(2796)/{\rm \AA}\,\rangle=-0.050\pm 0.25$.

Interpreting the scatter as due to Poisson noise in the number of
individual clumps intercepted along a line of sight leads to an
estimate of the mean number of clumps per galactic halo per sightline,
$n^{\rm clump}$ \citep[e.g.][]{Lanzetta:1990, Chen:2010}. In this
simple toy model, each absorber is characterized as $W_r=n^{\rm
  clump}\times \omega_0$, where $\omega_0$ is the mean absorption
equivalent width per component.  Following Poisson counting
statistics, the intrinsic scatter $\delta(\log\,W_r)\equiv
\delta(W_r)/(W_r\,\ln\,10)$ is related to the mean number of clumps
$\langle\,n^{\rm clump}\,\rangle$ according to
\begin{equation}
\delta(\log\,W_r) = \frac{1}{\ln\,10}\frac{\sqrt{\langle\,n^{\rm clump}\,\rangle+1}}{\langle\,n^{\rm clump}\,\rangle},
\end{equation}
which leads to $\langle\,n^{\rm clump}\,\rangle \sim 3.8$ for
$\langle\,W_r(2796)\,\rangle\approx 1$ \AA\ in the LRG halos with a
corresponding mean absorption equivalent width per component of
$\omega_0=0.24$ \AA.  The inferred $\langle\,n^{\rm clump}\,\rangle$
per sightline for LRG halos is more than four times smaller than what
was inferred for lower-mass, star-forming galaxies at $d\approx 40$
kpc \citep{Chen:2010}.  With a halo size twice as big as $L_*$
galaxies, the significantly lower $\langle\,n^{\rm clump}\,\rangle$
therefore suggests that the volume filling factor of \MgII\ absorbing
gas is $\sim 10$ times lower in LRG halos than in $L_*$ halos.

\section{Kinematics}
\label{sec:kinematics}

While obtaining spatially-resolved velocity maps is challenging for
both stellar and gaseous components of individual high-redshift
galaxies, important insights into the physical nature of cool halo
clouds can be gained from comparing the distribution of relative
velocities between absorbers and their associated galaxies.  One
important observable of cool halo gas is the line-of-sight velocity
dispersion within the host halo.  Using a sample of galaxy and
absorber pairs, it is possible to constrain the ensemble average of
the velocity distribution of absorbing gas relative to their host
galaxies.

\begin{figure}[!ht]
\begin{center}
\includegraphics[width=4.5in]{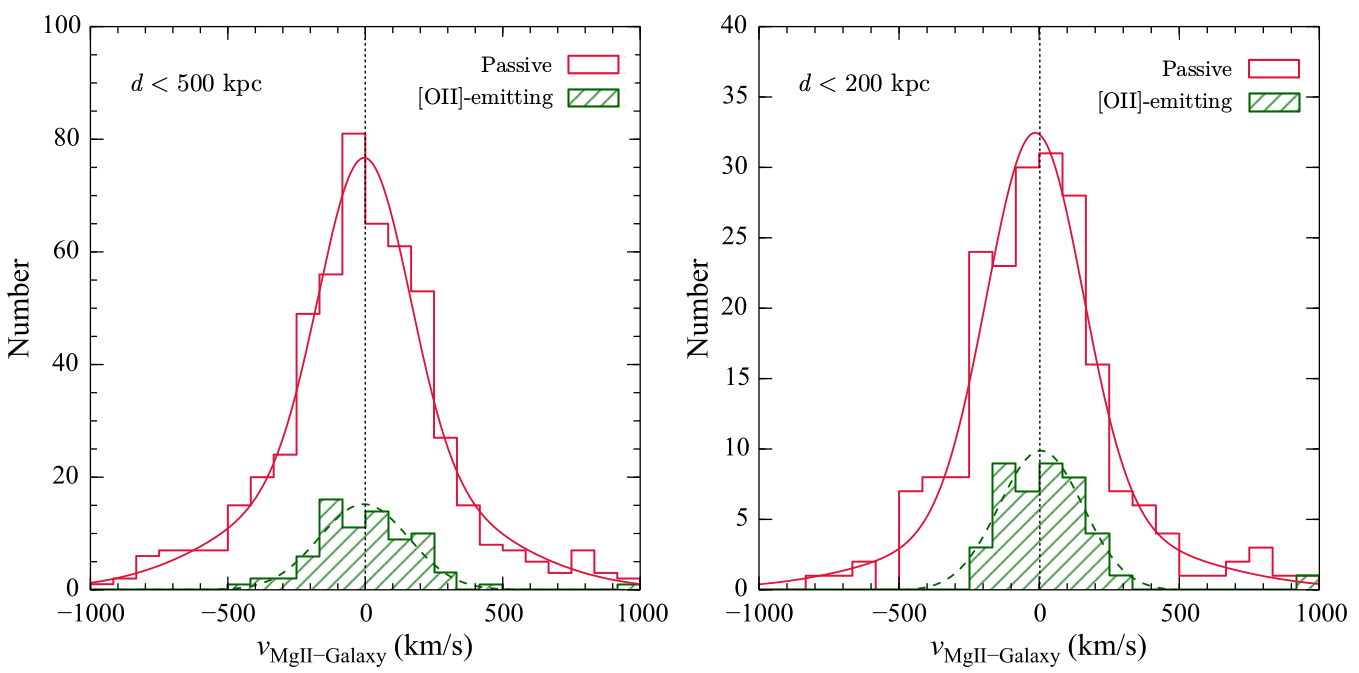}
\end{center}
%
%
\caption{Velocity distributions of \MgII-absorbing clumps relative to
  the systemic redshifts of LRGs based on an ensemble of \MgII\
  absorber and LRG pairs \citep[adapted from][]{Huang:2016}. The
  relative velocity distribution of \MgII-absorbing gas relative to
  passive LRGs is shown in the red open histogram, while the velocity
  distribution of \MgII\ absorbing gas relative to \OII-emitting LRGs
  is shown in the green hatched histogram.  Including all \MgII
  absorbers at $d<500$ kpc of an LRG ({\it left} panel), the velocity
  distribution of the gas is well characterized by a single Gaussian
  distribution centered at $\langle v_{\rm
    Mg\,\scriptsize{II}-Galaxy}\rangle = -5$ \kms\ with a dispersion
  of $\sigma_v = 167$ \kms\ (green, dashed curve).  At the same time,
  a double Gaussian profile is required to better characterize the
  velocity distribution of \MgII\ absorbing gas around passive LRGs
  (red solid curve) with a narrow component centered at $\langle
  v_{\rm Mg\,\scriptsize{II}-Galaxy}\rangle = -3$ \kms\ and $\sigma_v^n
  = 163$ \kms\ and a broad component centered at $\langle v_{\rm
    Mg\,\scriptsize{II}-Galaxy}\rangle = -17$ \kms\ and $\sigma_v^b =
  415$ \kms.  Roughly 12\% of passive LRGs have a \MgII\ absorber
  found at velocities exceeding 500 \kms\ from the systemic redshifts
  of the galaxies.  Consistent velocity distributions are found for
  \MgII\ absorbers detected at $d<200$ kpc from passive and
  [O\,II]-emitting galaxies ({\it right} panel), showing little
  variation in the overall gas motion between inner and outer halos.}
\label{fig:vdiff}       
\end{figure}

The distribution of relative velocity between \MgII\ absorbers and
their host galaxies is shown in Figure \ref{fig:vdiff} for both
passive and [O\,II]-emitting LRGs.  For the two LRG subsamples, the
relative motion of \MgII\ absorbing gas is well centered around the
systemic redshifts of the galaxies.  While the velocity distribution
around [O\,II]-emitting galaxies is well characterized by a single
Gaussian of dispersion $\sigma_v =167$ \kms, a double Gaussian profile
is required to characterize the relative motion of \MgII\ gas around
passive LRGs with a narrow component of $\sigma_v^n =163$ \kms\ for
the majority of the sample and a broad component of $\sigma_v^b = 415$
\kms.  Roughly $12$\% of passive LRGs are found in the broad component
with velocities exceeding 500 \kms. These properties apply to both
inner ($d<200$ kpc) and outer ($200<d<500$ kpc) halos
\citep[see][for a detailed discussion]{Huang:2016}.

The implications of Figure \ref{fig:vdiff} are two-fold.  First of
all, the similarity in the relative velocity distribution of \MgII\
absorbing gas around passive and [O\,II]-emitting LRGs reaffirms the
findings from the previous section (Section \ref{sec:spatial}) that
the observed \MgII\ absorbers are unlikely to originate in outgoing
materials from the LRGs.  Secondly, the mean halo mass of LRGs is
$M_h\approx 3\times 10^{13}\,M_\odot$ \citep[e.g.][]{Mandelbaum:2008,
  Gauthier:2009}. The expected line-of-sight velocity dispersion for
virialized gas in these massive halos is $\sigma_h\approx 265$ \kms.
The observed velocity dispersion of \MgII\ absorbing gas in $\approx
90$\% of the LRG sample (the narrow component in Figure
\ref{fig:vdiff}) is merely 60\% of what is expected from virial motion
\citep[see also][]{Zhu:2014}, namely
\begin{equation}
\sigma_v\approx 0.6\,\sigma_h.
\end{equation}  
Such suppression in gas motion not only indicates that the gas is
gravitationally bound in the LRG halo but also that the kinetic energy
of the gas is being dissipated.  Similar suppression in the
line-of-sight velocity dispersion is also found for \lya\ and O\,VI
absorbing gas around COS-Halos red galaxies with $M_{\rm star}\apg
10^{10.7}\,\msol$ \citep{Tumlinson:2011, Tumlinson:2013}.

In the presence of a hot halo \citep[e.g.][]{Hand:2011}, cool clumps
should experience a ram-pressure drag force and slow down.  For
ram-pressure to drive decelerations in the clump motion, these clumps
cannot be too massive.  The observed suppression in the velocity
dispersion of the absorbing gas therefore places a mass limit on these
clumps.  Following the formalism of \cite{Maller:2004}, 
a maximum clump mass can be estimated for the
LRG halos according to $m_{\rm cl}\approx 5.1\times
10^4\,M_\odot\,T_6^{-3/8}(\Lambda_z\,t_8)^{1/2}$, where $T_6$ is the
halo gas temperature in units of $10^6\, \rm K$, $\Lambda_z$ is the
cooling parameter that varies with the gas metallicity $\rm Z_g$ and
$t_8 = t_f(c_h)/8\, {\rm Gyr}$ is the halo formation time that depends
on the halo concentration $c_h$.  For a halo temperature of $T\sim
6\times10^6\,\rm K$ for the LRGs and $t_8 \sim 8.9$ Gyr using
$c_h\sim10$ from the halo mass--concentration relation
\citep[e.g.][]{Mandelbaum:2008}, the estimated mass of individual \MgII\
absorbing clumps is $m_{\rm cl}\approx 5\times 10^4\,M_\odot$ for
solar metallicity and lower for lower-metallicity
\citep[see][for more details]{Huang:2016}.

Beyond an ensemble average over a large sample of LRG-\MgII\ pairs,
comparisons of gas and satellite kinematics in individual halos are
possible when deep galaxy survey data are available
\citep[e.g.][]{Steidel:1997, Whiting:2006, Nestor:2007, Chen:2009}.  Previous
studies that combined absorber and galaxy data have revealed a number
of metal-line absorbers associated with overdensities of galaxies over
a wide range of mass scales \citep[e.g.][]{Mulchaey:2009, Kacprzak:2010,
  Nestor:2011, Gauthier:2013, Johnson:2013}.  Comparing the observed
line-of-sight velocity dispersion of the absorbing gas with the
velocity distribution of galaxy group members offers additional clues
for the origin of chemically-enriched gas in overdense environments.
In particular, imaging and spectroscopic surveys of galaxies in the
vicinities of ultra-strong \MgII\ absorbers, $W_r(2796)\apg 3$ \AA,
have revealed multiple galaxies at small projected distances and
velocity separations from these absorbers \citep[e.g.][]{Nestor:2007,
  Nestor:2011, Gauthier:2013}.  With the expected velocity spread of
$\sim\,100$ \kms\ per Angstrom for \MgII\ absorbers, the large
$W_r(2796)$ implies a line-of-sight velocity spread of the components
$\Delta\,v > 300$ \kms.  Attributing the observed gas motion to the
underlying gravitational potential of the host dark matter halo would
lead to a halo mass exceeding $M_h\approx 3\times 10^{12}\,\msol$.

\begin{figure}[!th]
\begin{center}
\includegraphics[width=4.5in]{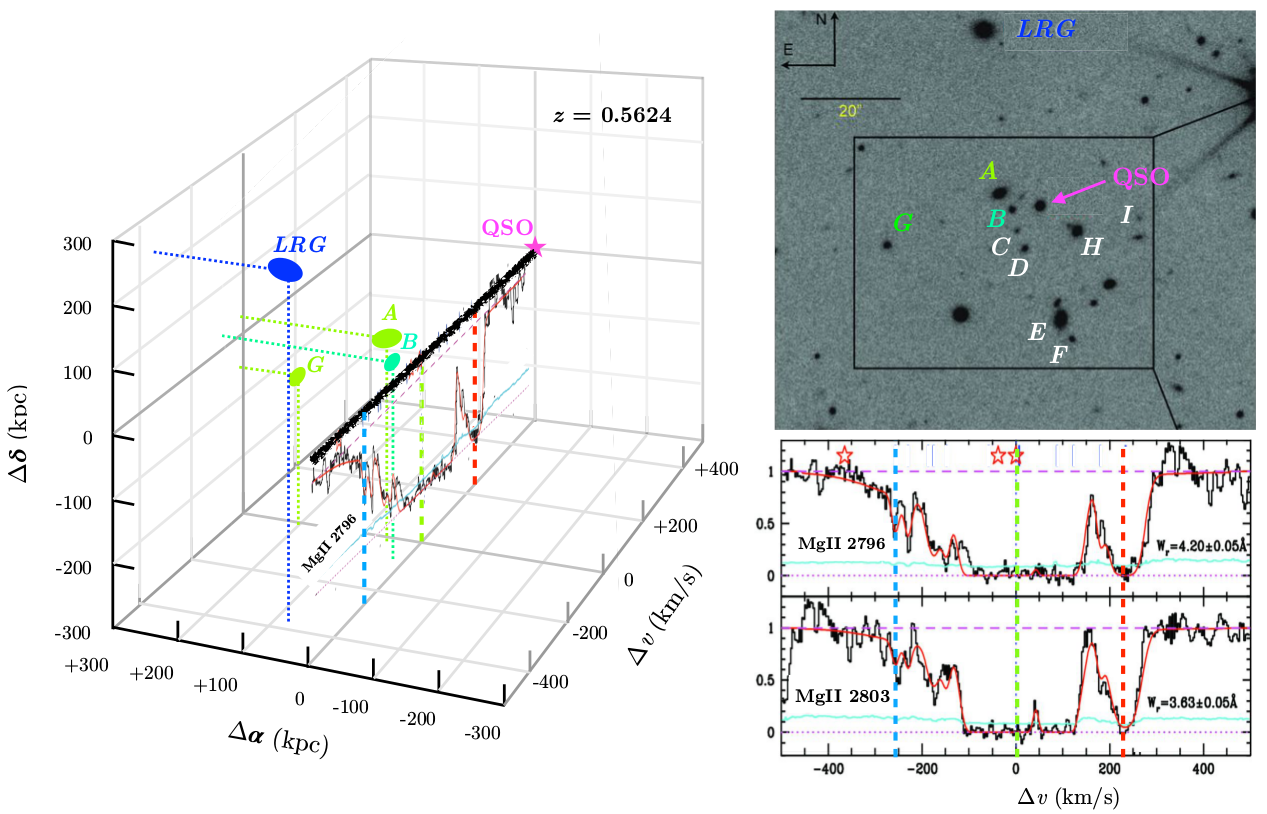}
\end{center}
\caption{Comparison of galaxy and \MgII\ absorption gas kinematics
  observed in a group of galaxies at $z=0.5624$
  \citep{Gauthier:2013}.  The imaging panel and absorption-line
  profiles on the right are adapted from
  \cite{Gauthier:2013}.  An ultra-strong \MgII\ absorber of
  $W_r(2796)=4.2$ \AA\ ({\it lower-right} panel) is identified at
  $z=0.5624$ along the sightline toward a background QSO at $z_{\rm
    QSO}=1.30$.  Follow-up spectroscopy of galaxies in this field
  ({\it upper-right} panel) revealed an LRG at $d=246$ kpc and
  $\Delta\,v=-385$ \kms, along with three additional galaxies, $A$,
  $B$, and $G$ at closer distances and relative velocities from the
  \MgII\ absorber.  Galaxies $C-F$, $H$, and $I$ are found to be at
  lower redshifts from the galaxy group at $z=0.56$.  The left panel
  displays the spatial and relative velocity distribution of the LRG
  and three neighboring galaxies relative to the \MgII\ absorber.  The
  luminosities of the LRG and galaxies $A$, $B$, and $G$ are 3.5, 1.8,
  0.3, and $0.5\,L_*$, respectively.  The galaxies are color-coded
  according to the relative velocities from the \MgII\ absorber at
  $\Delta\,v=0$ \kms.  Galaxies $A$ and $G$, marked in pale green,
  occur at $\Delta\,v\approx 0$ \kms, while the LRG, marked in blue,
  displays the largest blue-shifted relative velocity with
  $\Delta\,v=-385$ \kms.  The most blue-shifted and redshifted \MgII\
  components at $\Delta\,v\approx -250$ and 230 \kms\ are also marked
  in pale blue and red, respectively, to guide the visual comparison.}
\label{fig:group}       
\end{figure}

Figure \ref{fig:group} showcases an example of an ultra-strong \MgII\
absorber of $W_r(2796)=4.2$ \AA\ identified at $z=0.5624$ along the
sightline toward a background QSO at $z_{\rm QSO}=1.30$
\citep{Gauthier:2013}.  Four galaxies are spectroscopically
identified at close projected distances and velocity separations from
the absorber, including a 3.5-$L_*$ LRG at $d=246$ kpc and
$\Delta\,v=-385$ \kms, a 1.8-$L_*$ passive galaxy ($A$) at $d=55$ kpc
and $\Delta\,v\approx 0$ \kms, a 0.3-$L_*$ galaxy ($B$) at $d=38$ kpc
and $\Delta\,v\approx -100$ \kms, and a 0.5-$L_*$ galaxy ($G$) at
$d=209$ kpc and $\Delta\,v\approx 0$ \kms.  Only galaxies $B$ and $G$
exhibit traces of on-going star formation with an inferred star
formation rate of $\approx 3\,\msol\,{\rm yr}^{-1}$.  The total
stellar masses of the galaxies range from $M_{\rm star}\apg 4\times
10^{10}\,\msol$ for galaxy $A$ and the LRG to $M_{\rm star}\approx
3\times 10^{9}\,\msol$ for galaxies $B$ and $G$.  The close proximity
and the observed massive stellar content qualify these galaxies as
parts of a galaxy group with a light-weighted center at $d_{\rm
  group}\approx 156$ kpc and $\Delta\,v_{\rm group}= -200$ \kms.  The
total dark matter halo mass of the group estimated based on the
stellar mass of the LRG is $M_h\approx 10^{13}\,\msol$
\cite[e.g.][]{Behroozi:2010}.

The star formation rate intensity observed in galaxies $B$ and $G$ is
small \citep{Gauthier:2013}, which indicates that super galactic
winds are unlikely to form and drive outflows in these galaxies
\citep[e.g.][]{Heckman:2002, Heckman:2015}.  On the other hand, the large
line-of-sight velocity spread of the absorbing gas, from
$\Delta\,v\approx -250$ to $\Delta\,v\approx 230$ \kms\ (lower-right
panel), is comparable to the projected velocity dispersion expected
from virial motion, $\sigma_{\rm vir}\approx 190$ \kms, in a halo of
$M_h\approx 10^{13}\,\msol$ at $z=0.5$.  This agreement suggests a
physical connection between group dynamics and ultra-strong \MgII\
absorbers.  A comparable velocity dispersion between absorbing gas and
group galaxies has also been found for a \MgII\ absorber of
$W_r(2796)=1.8$ \AA\ at $z=0.3$, for which a group of five sub-$L_*$
galaxies are found at $d<250$ kpc and $|\Delta\,v|<250$
\kms\ \citep{Kacprzak:2010}.  The dynamical mass of the host dark
matter halo, based on the velocity dispersion of the group members
$\sigma_v=115$ \kms, is $M_h\approx 3\times 10^{12}\,\msol$ for the
galaxy group at $z=0.3$.  In both cases, where a strong \MgII is found
in a galaxy group, dissipation is not observed in the relative motion
between the absorbing gas and associated galaxies.  The large velocity
dispersion, similar to the broad velocity component seen in Figure
\ref{fig:vdiff}, can be understood if the ram-pressure drag force is
subdominant in comparison to the gravitational force on the gas
clumps.  Stripped gas due to tidal interactions between neighboring
galaxies is a likely explanation for the dynamically complex \MgII\
absorbers found in group or cluster environments
\citep[e.g.][]{Chynoweth:2008, Marasco:2016}, which also results in H\,I
depleted galaxies in these massive cluster and group halos
\citep[e.g.][]{Verdes:2001, Chung:2009}.

Galaxy surveys have been carried out for two more ultra-strong \MgII\
absorbers at $z\approx 0.7$.  Both are found in a group environment
with $M_h=10^{12-13}\,\msol$ \citep{Nestor:2011}.  However,
high-resolution spectra are not available for these two absorbers and
consequently comparisons of gas and galaxy kinematics are not
possible.


\section{Chemical Enrichment}
\label{sec:metal}

Figure \ref{fig:spatial} clearly indicates that despite being in a
quiescent state, halo gas around massive LRGs has been enriched with
heavy elements out to the virial radius.  The chemical enrichment
history in these massive halos, which can be broadly characterized by
(i) gas metallicity and (ii) relative abundance pattern, offers
independent clues for the origin of the observed cool halo gas.
Specifically, gas metallicity quantifies the overall heavy element
production level, and nearby galaxies are observed to follow a
mass--metallicity relation with more massive galaxies displaying on
average higher metallicities both in stars \citep{Gallazzi:2005,
  Kirby:2013} and in the ISM \citep{Tremonti:2004, Lee:2006}.
Therefore, a natural expectation is that gas ejected from massive
galaxies should be highly metal-enriched, while gas ejected from
low-mass satellites should exhibit a lower metallicity.  Gas accreted
from the IGM should contain still lower metallicities 
\citep[e.g.][]{Lehner:2013, Kacprzak:2014}.  However, uncertainties arise as a
result of poorly understood processes involving mixing and transport
of heavy elements \citep[e.g.][]{Tumlinson:2006}.  If mixing is
effective as metal-enriched materials propagate into the low-density,
and presumably lower-metallicity, halo gas, then the overall
metallicity is expected to be reduced.  If mixing is ineffective, then
large variations in chemical abundances are expected in single objects
\citep[e.g.][]{Scalo:2004}.  Consequently, metallicity provides only a
weak indicator for possible origins of chemically-enriched cool gas in
LRG halos.

On the other hand, the relative abundance pattern presents a fossil
record for the primary source of heavy element production.
Specifically, $\alpha$-elements (such as O, Mg, Si, and S) are
primarily produced in massive stars and core-collapse supernovae
(SNe), while a significant fraction of heavier elements such as Fe and
Ni are produced in Type Ia SNe over longer timescales
\citep[e.g.][]{Tsujimoto:1995, Nomoto:2006}.  The observed [$\alpha$/Fe]
relative abundance ratio therefore constrains the relative
contributions of core-collapse SNe and SNe\,Ia to the chemical
enrichment history \citep{Tinsley:1979, Gilmore:1991, Tolstoy:2003}.
Both distant galaxies \citep{Wolfe:2005} and nearby evolved stars
\citep{Matteucci:2014} display an $\alpha$-element enhanced abundance
pattern, indicating a chemical enrichment process driven by
core-collapse SNe in the early universe.  At the same time, the ISM of
nearby elliptical galaxies \citep[e.g.][]{Mathews:2003, Humphrey:2006}
and intracluster medium \citep[e.g.][]{dePlaa:2007} display super-solar
Fe/$\alpha$ abundance ratios that reveal increased contributions from
SNe Ia in the circumgalactic gas of massive, evolved galaxies.
Relative abundance measurements between Fe and $\alpha$-elements
therefore provide a powerful constraint for the origin of
chemically-enriched gas around galaxies.

\begin{figure}[!th]
\begin{center}
\includegraphics[width=4.5in]{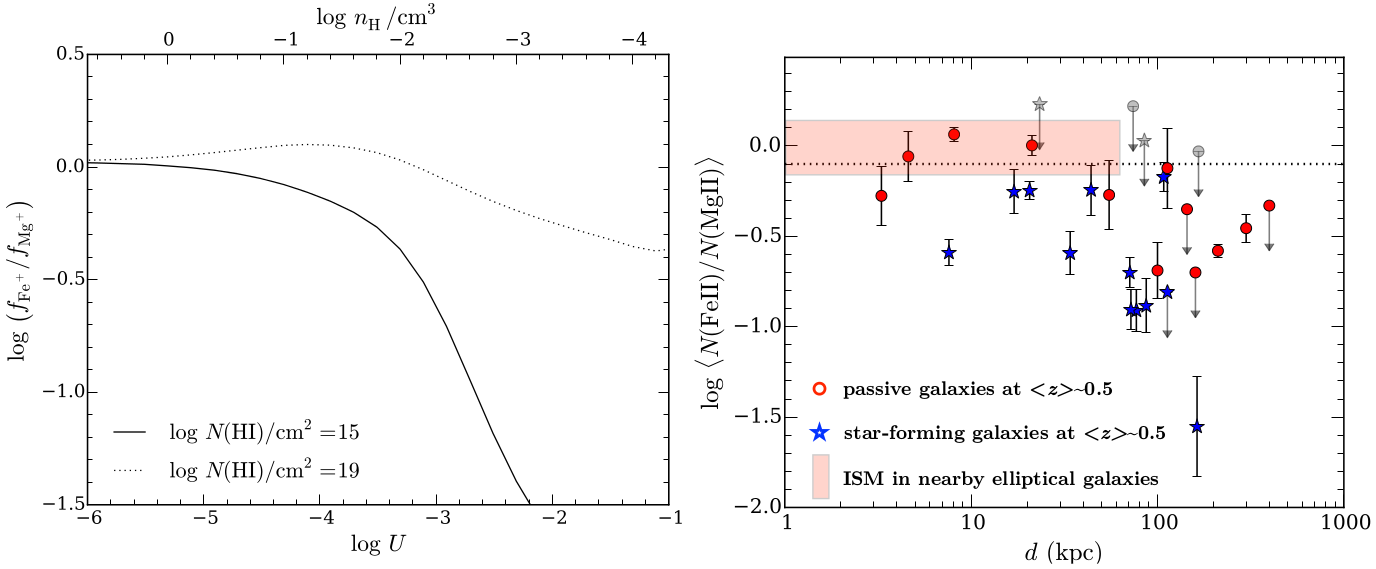}
\end{center}
\caption{Enhanced Fe abundances in halos around massive quiescent
  galaxies in comparison to those around star-forming galaxies
  \citep{Zahedy:2016a, Zahedy:2016b}.  {\it Left}: Expected ionization fractions of
  Mg$^+$ and Fe$^+$ ions from Cloudy photo-ionization calculations
  \citep{Ferland:2013} for photo-ionized gas of temperature $T=10^4$
  K and 0.1 solar metallicity in optically-thin (with neutral hydrogen
  column density $N({\rm H\,I})=10^{15}\,\cmjj$) and optically-thick
  (with $N({\rm H\,I})=10^{19}\,\cmjj$) regimes \citep{Zahedy:2016a}.
  Changing metallicity does not alter the results.  An updated version
  of the \cite{Haardt:2001} ionizing radiation field
  at $z=0.5$ is adopted for computing the ionization parameter $U$ at
  different gas density $n_{\rm H}$ displayed at the top of the panel.
  These calculations demonstrate that, independent of gas metallicity,
  the observed $N({\rm Fe\,II})/N({\rm Mg\,II})$ in the right panel
  represents a lower limit to the underlying $({\rm Fe}/{\rm Mg})$
  relative abundances.  {\it Right}: Observed mean Fe\,II to \MgII\
  column density ratio versus projected distance based on absorption
  spectroscopy of halo gas around quiescent (red circles) and
  star-forming (blue stars) galaxies at $\langle\,z\,\rangle\approx
  0.5$.  For comparison, ISM measurements based on x-ray observations
  of 19 nearby elliptical galaxies \citep{Humphrey:2006} are also
  included in the left panel, and the dotted line shows the solar
  value.  The fractional contribution of Type Ia supernovae to the
  chemical enrichment history in the inner 60 kpc of massive quiescent
  halos is found to be similar to what is observed in the solar
  neighborhood.}
\label{fig:composition}       
\end{figure}

At $z\apg 0.3$, a series of Fe\,II absorption transitions are
observable on the ground, which enable an accurate measurement of
Fe\,II column density $N({\rm Fe\,II})$ on a component by component
basis \citep[e.g.][]{Rigby:2002, Narayanan:2008, Zahedy:2016a}.
Similarly, the \MgII\ doublet transitions enable accurate \MgII\
column density measurements $N$(\MgII) for all but a few
saturated components.  In principle, the total elemental abundances
can be determined from the observed $N$(\FeII) and $N${\MgII), 
if the ionization fractions of these ions are known.  Gas
metallicity can also be derived if the total hydrogen content can be
inferred from measurements of the neutral hydrogen column density,
$N({\rm H\,I})$.  In practice, observations of $N({\rm H\,I})$ at low
redshifts require UV spectroscopy carried out in space, and have only
been done for a relatively small sample of galaxies
\citep[e.g.][]{Chen:2000, Werk:2014}.  Furthermore, uncertainties in the
derived ionization fractions of individual ions can be large and
difficult to estimate under simple ionization models.

Observations of the relative abundance ratio between Mg$^+$ and Fe$^+$
ions are particularly useful, because Mg$^+$ and Fe$^+$ share similar
ionization potentials (15 eV and 16.2 eV, respectively) and are the
dominant ionization states of the respective elements in both neutral
and cool photo-ionized gas.  While the elemental abundances of iron
and magnesium may be uncertain, the relative abundance ratio, $({\rm
  Fe}/{\rm Mg})$ can be inferred with high confidence from the
observed $N({\rm Fe\,II})/N({\rm Mg\,II}$ following $\log\,\left({\rm
  Fe}/{\rm Mg}\right)=\log\,N({\rm Fe\,II})/N({\rm Mg\,II}) -
\log\,(f_{{\rm Fe}^+}/f_{{\rm Mg}^+})$, where $f_{{\rm Fe}^+}$ is the
fraction of Fe in the singly ionized state and $f_{{\rm Mg}^+}$ is the
fraction of Mg in the singly ionized state.  The ratio of ionization
fractions, $f_{{\rm Fe}^+}/f_{{\rm Mg}^+}$, estimated based on a suite
of photo-ionization calculations is shown in the left panel of Figure
\ref{fig:composition}.  The photo-ionization models are computed using
the Cloudy software \citep{Ferland:2013}.  These models assume a
photo-ionized gas of temperature $T=10^4$ K and 0.1 solar metallicity
in optically-thin (with neutral hydrogen column density $N({\rm
  H\,I})=10^{15}\,\cmjj$) and optically-thick (with $N({\rm
  H\,I})=10^{19}\,\cmjj$) regimes \citep{Zahedy:2016a}.  An updated
version of the \cite{Haardt:2001} ionizing radiation
field at $z=0.5$ is adopted for computing the ionization parameter $U$
(defined as the number of ionizing photons per atom) at different gas
density $n_{\rm H}$ displayed at the top of the panel.  The ionization
fraction of Fe$^+$ remains roughly equal to that of Mg$^+$ in the
optically-thick regime and lower in optically-thin gas for the full
range of $U$ explored.  Consequently, $\log\,(f_{{\rm Fe}^+}/f_{{\rm
    Mg}^+}) \apl 0$ and the observed $N({\rm Fe\,II})/N({\rm Mg\,II})$
marks a lower limit to the underlying Fe to $\alpha$-element abundance
ratio:
\begin{equation}
\log\,\left(\frac{{\rm Fe}}{{\rm Mg}}\right)>\log\,\frac{N({\rm Fe\,II})}{N({\rm Mg\,II})}.
\end{equation}
Experimenting with different gas metallicity does not change the
predicted ionization fraction, and accounting for differential dust
depletion of iron and magnesium would further increase the inferred
$[{\rm Fe}/{\rm Mg}]$ \citep{Savage:1996}.  This exercise
demonstrates that useful empirical constraints for the relative Fe and
Mg abundances can be obtained even in the absence of accurate
measurements of $N({\rm H\,I})$ and gas metallicity.

A recent study utilizing multiply-lensed QSOs for probing gas at small
projected distances, $d\apl 20$ kpc or $\approx 1-2$ effective radii
$r_e$, from the lensing galaxies has uncovered important new clues for
the origin of chemically-enriched cool gas in massive halos
\citep{Zahedy:2016a}.  These lensing galaxies at $z=0.4-0.7$ share
similar properties concerning both the quiescent state and halo mass
scales ($M_h\apg 10^{13}\,\msol$), but display distinct
absorption-line profiles between different lensing galaxies and
between different sightlines of individual lenses.  The apparent large
scatter in the observed absorption profiles is consistent with the
large scatter displayed in Figure \ref{fig:spatial} and discussed in
Section \ref{sec:spatial}.  Most interestingly, all \MgII\ absorbers
detected near these lensing galaxies are strong and resolved into
$8-15$ individual components over a line-of-sight velocity range of
$\Delta\,v \approx 300 - 600$ \kms.  The \MgII\ absorption is
accompanied with even stronger Fe\,II absorption with matching
kinematic profiles.  Comparing the relative absorption strengths
between individual components also yields uniformly large $N({\rm
  Fe\,II})/N({\rm Mg\,II})$ ratios over the full range of velocity
spread, $\Delta\,v$, with a median of $\langle\,\log\,N({\rm
  Fe\,II})/N({\rm Mg\,II})\,\rangle \approx 0$ and a scatter of $<0.1$
dex.

Following Equation (3), the observed $\langle\,\log\,N({\rm
  Fe\,II})/N({\rm Mg\,II})\,\rangle\approx 0$ naturally leads to a
super solar Fe/Mg abundance ratio near these massive lensing galaxies.
This is in stark contrast to an $\alpha$-element enhanced chemical
composition found in young, star-forming galaxies
\citep[e.g.]{Dessauges:2004, Crighton:2013, Fox:2014} and see \cite{Zahedy:2016a} for a detailed 
comparison, and clearly indicates
different origins between gas associated with star-forming galaxies
and with massive quiescent galaxies.

Previous studies have shown that the spatial distribution of SNe~Ia in
nearby early-type galaxies follows the stellar light out to $r\sim
4\,r_e$ \citep[e.g.][]{Forster:2008} and that the ISM of these massive
quiescent galaxies exhibits an iron-enhanced abundance pattern
\citep[e.g.][]{Mathews:2003, Humphrey:2006}.  Therefore, it seems likely that
the observed Fe-rich gas at $d \approx r_e$ from the lenses originates
in the ISM of these massive galaxies, where SNe~Ia play a dominant
role in driving the observed large velocity width and Fe-rich
abundance pattern \citep{Zahedy:2016a}.  If the ISM of massive
galaxies at intermediate redshift is locally enriched by SNe~Ia, then
the observed Fe/Mg is expected to decline with increasing projected
distance.  

To test this hypothesis, a sample of 13 massive quiescent galaxies
(including two lensing galaxies) and 14 star-forming galaxies at
intermediate redshifts ($\langle\,z\,\rangle\approx 0.5$) has been
assembled \citep{Zahedy:2016b}.  These galaxies were selected to have
absorption spectra of background QSOs at $d\approx 10-400$ kpc
available for constraining the radial profile of Fe/Mg.  The right
panel of Figure \ref{fig:composition} displays the mean
$\langle\,N({\rm Fe\,II})/N({\rm Mg II})\rangle$ averaged over all
individual components per halo versus $d$ for both star-forming
galaxies (blue stars) and the quiescent galaxy population (red
circles).  Although the sample is still small, the distinction between
quiescent and star-forming halos is already apparent in Figure
\ref{fig:composition}.  While the mean Fe/Mg ratio is consistent with
an $\alpha$-element enhanced pattern in the outer halo at $d>100$ kpc
for both quiescent and star-forming galaxies, halo gas at $d<100$ kpc
from quiescent galaxies exhibits an elevated iron abundance in
comparison to star-forming galaxies.

The observed iron enrichment level in the inner halo of $z\approx 0.5$
quiescent galaxies is consistent with the solar value, similar to what
is observed in the ISM of nearby elliptical galaxies 
\citep[e.g.][]{Mathews:2003, Humphrey:2006} and in the intracluster medium
\citep[e.g.][]{dePlaa:2007}.  The radial decline of the Fe/Mg relative
abundances supports the hypothesis that the gas is locally enriched by
SNe~Ia.  The minimum fractional contribution of SNe~Ia to the chemical
enrichment in the inner halos of massive quiescent galaxies is found
to be $f_{\rm Ia}\approx 15-20$\% based on the expected yields for
Type Ia and core-collapse SNe \citep{Iwamoto:1999} and the observed
Fe/Mg ratio at $d<100$ kpc.

\section{Quasar Host Halos}
\label{sec:qso}

In comparison to LRGs which are massive and quiescent, quasars powered
by supermassive black holes also reside in massive halos of 
$M_h\apg 3\times 10^{12}\,\msol$ \citep{Porciani:2004, White:2012, Shen:2013Q}
but in an active phase that lasts $\approx 10-100$ Myr
\citep{Martini:2004}.  While the physical processes that drive the
fueling and feedback of the central supermassive black holes are not
well understood \citep[e.g.][]{Hopkins:2009, Heckman:2014}, quasar
feedback is a critical ingredient in shutting down star formation in
high-mass halos and producing red-and-dead elliptical galaxies in all
theoretical frameworks \citep[e.g.][]{Benson:2003, Croton:2006}.
Indeed, most nearby early-type galaxies are found to harbor a
supermassive black hole at the center \citep{Kormendy:1995,Ho:2008}.
Therefore, studying quasar hosts is an integral part of the effort to
understand the growth of massive galaxies and observations of quasar
host halos provide important clues for how the halo gas properties are
shaped while the galaxies undergo an active quasar phase.

\begin{figure}[!ht]
\begin{center}
\includegraphics[width=4.5in]{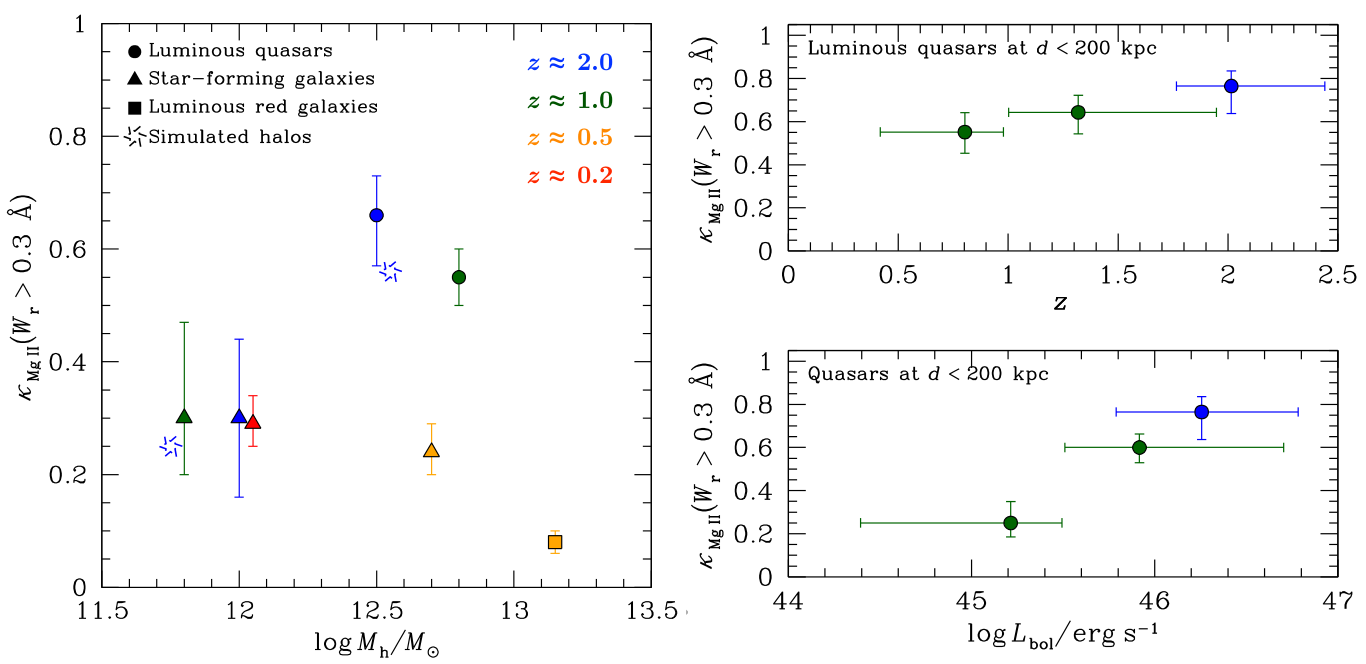}
\end{center}
\caption{Elevated cool gas covering fraction in quasar host halos.
  {\it Left}: Mean gas covering fraction $\kappa$ as a function of
  halo mass for galaxies and quasar hosts at different redshifts.  The
  measurements for galaxies and quasars at $z<2$ are based on surveys
  of \MgII\ absorbers with $W_r({\rm Mg\,II})>0.3$ \AA\ and at $d <
  R_{\rm vir}$, where $R_{\rm vir}$ represents the virial radius of
  the host dark matter halos.  Measurements for $z\approx 2$
  star-forming galaxies \citep{Rudie:2012} and quasars
  \citep{Prochaska:2014} are inferred from the observed incidence of
  optically-thick H\,I absorbers and C\,II absorption observations,
  respectively.  Closed triangles represent star-forming galaxies
  \citep{Chen:2010, Lovegrove:2011, Rudie:2012, Werk:2014}, squares
  represent quiescent LRGs \citep{Huang:2016}, and circles represent
  quasar hosts \citep{Prochaska:2014,Johnson:2015}.  Star symbols are
  simulation predictions for $z\approx 2$ objects
  \citep{Faucher:2016}.  Different colors indicate different redshift
  ranges of each dataset with red, orange, green, and blue
  corresponding to $z \approx 0.2$, 0.5, 1.0, and 2.0, respectively.
  The mean gas covering fraction in quasar host halos is observed to
  be more than doubled from what is seen in star-forming galaxies at
  both low and high redshifts.  Numerical simulations have attributed
  the elevated cool gas content in quasar hosts to either AGN feedback
  \citep{Rahmati:2015} or star formation feedback from neighboring
  satellite galaxies \citep{Faucher:2016}.  {\it Right}: Mean gas
  covering fraction at $d < 200$ kpc for quasar hosts at different
  redshifts (top) and with different bolometric luminosity (bottom).
  Colors denote different redshift ranges as defined in the left
  panel.  While the mean gas covering fraction in quasar hosts is
  found to evolve only weakly with redshift, a steep correlation is
  observed between cool gas covering fraction and quasar luminosity.
  The elevated cool halo gas covering fraction observed in quasar
  hosts appears to be driven primarily by the most luminous quasar
  population.  (Figure credit: Sean Johnson)}
\label{fig:qso}       
\end{figure}

It has long been recognized that the intense radiation field from
luminous quasars is sufficient to fully ionize their surrounding
medium and to keep the IGM ionized over most of the cosmic history of
the universe.  Near the quasar redshifts, the incidence of
\lya\ absorption lines in the quasar spectra is reduced due to an
enhanced radiation field from the quasars in the proximity zone
\citep[e.g.][]{Bajtlik:1988}.  The line-of-sight quasar proximity effect,
which is also observed in different ionic transitions such as \MgII\
and C\,IV, offers a unique means of constraining both the metagalactic
ionizing radiation and the extent of the ionizing bubble local to the
quasar \citep{Dallaglio:2008, Wild:2008}.  However, a reduced
incidence of absorption features is not observed along the {\it
  transverse} direction from quasars \citep[e.g.][]{Fernandez:1995,
  Schirber:2004}.  In addition, there also appears to be an excess of
cool clouds in quasar halos \citep{Bowen:2006, Hennawi:2007,
  Farina:2013, Farina:2014, Prochaska:2014}.

The left panel of Figure \ref{fig:qso} shows the mean covering
fraction of cool halo gas $\kappa$ as a function of halo mass for
galaxies (closed triangles and squares) and quasars (closed circles)
at different redshifts.  For comparison, expectations from numerical
simulations are also included (open star symbols).  Observations of
halos around galaxies and quasars at $z<2$ are based on surveys of
\MgII\ absorbers with $W_r({\rm Mg\,II})>0.3$ \AA\ at projected
distances less than the virial radius $R_{\rm vir}$ of the host dark
matter halos.  Observations of $z\approx 2$ star-forming galaxies
\citep{Rudie:2012} and quasars \citep{Prochaska:2014} are inferred
from the observed incidence of optically-thick H\,I absorbers and
C\,II absorption features, respectively.  While halos around normal
galaxies display a roughly constant cool gas covering fraction of
$\kappa\approx 30$\% for the star-forming population at all redshifts
studied \citep[e.g.][]{Chen:2010, Lovegrove:2011, Rudie:2012} and a
suppressed cool gas content with $\kappa\apl 10$\% for the quiescent
population at $z\approx 0.5$ \citep[e.g.][]{Gauthier:2011, Huang:2016},
there appears to be a surge of cool gas with $\kappa\approx 60$\% in
active quasar halos \citep{Prochaska:2014, Farina:2013, Farina:2014,
  Johnson:2015Q} at both low and high redshifts.

While state-of-the-art numerical simulations can reproduce the
observed incidence of cool halo gas around star-forming galaxies and
quasars at $z\approx 2$, the underlying physical mechanisms are
fundamentally different.  The elevated cool gas content in quasar
hosts are attributed to either AGN feedback \citep{Rahmati:2015} or
star formation feedback from neighboring satellite galaxies
\citep{Faucher:2016}.

The observed line-of-sight proximity effect, together with an absence
of transverse proximity effect, indicates that the quasar radiation
field is not isotropic and that cool clouds outside of the radiation
zone can still survive in the host halos.  Understanding whether and
how the incidence of cool gas varies with quasar properties provides
further insights into the origin of the observed excess of cool halo
gas around quasars.  Using a sample of 195 projected quasar pairs
separated by less than 300 kpc in projected distance, it has been
shown that the mean gas covering fraction (measured based on
observations of \MgII\ absorption transitions) correlates strongly
with the bolometric luminosity $L_{\rm bol}$ of the foreground quasar
\citep{Johnson:2015Q}.  Specifically, quasars that are ten times more
luminous display on average $3-4$ times higher cool gas covering
fraction than low-luminosity quasars at $d<200$ kpc in the halos
(lower-right panel of Figure \ref{fig:qso}).  The elevated cool halo
gas covering fraction observed in quasar hosts appears to be driven
primarily by the most luminous quasars in the sample.

The observed strong correlation between $\kappa$ and $L_{\rm bol}$
is unlikely to be driven by an underlying mass dependence
\citep[c.f.][]{Chen:2010m}, because the clustering amplitude of quasars
remains roughly constant in different luminosity intervals
\citep{Shen:2013Q, Eftekharzadeh:2015}.  The observed $L_{\rm bol}$
dependence in cool halo gas covering fraction therefore has profound
implications for the connection between halo gas on the 100 kpc scale
and quasar activities in the central parsec scale.

Possible explanations for the observed excess of cool gas along
transverse sightlines include: (i) overdensity of galaxies in the
quasar environment which is expected from the large clustering
amplitude observed for luminous quasars, (ii) outflows from the
quasar/nuclear starburst, and (iii) debris from galaxy interactions or
mergers that trigger the luminous quasar phase.  Of these scenarios,
quasar outflows offer the most promising explanation but caveats
remain.  Significant contributions from correlated galaxies sharing
the same large-scale overdensity peak can be ruled out based on the
diminishing $\kappa$ observed at $d \apg 200$ kpc from quasars
\citep{Johnson:2015Q}.  Merger remnants are unlikely to explain the
observed cool gas at $d\sim 100$ kpc, given that the quasar lifetime
is shorter than the dynamical time.

Spatially extended outflows have been detected in [O\,III] emission
out to $10-30$ kpc around luminous quasars at $z\sim 0.5$ with outflow
velocities as high as $v_{\rm out}\approx 1000$ \kms\
\citep[e.g.][]{Greene:2012, Liu:2013, Liu:2014}.  The observed spherical
morphologies in these [O\,III] emitting nebulae suggest that outflows
in these quasar hosts are not well-collimated.  It is therefore
possible that the absorbers detected at $d\sim 100$ kpc along
transverse sightlines in quasar halos originate in these extended
outflows with densities too low to be detected in emission.  

An outflow origin also provides a natural explanation for the observed
strong correlation between $\kappa$ and $L_{\rm bol}$ with more
luminous quasars driving higher mass outflow rates
\citep[e.g.][]{Carniani:2015}.  This is at least qualitatively consistent with
the extreme kinematics displayed in some ($\approx 10$\%) quasar
absorbers that spread over a velocity interval of 
$|\Delta\,v|\apg 1000$ \kms\ \citep[e.g.][]{Johnson:2015Q}. 
However, free-expanding
outflows, traveling at $v_{\rm out} = 1000$ \kms\ at 15 kpc from the
quasar, are expected to reach 100 kpc in $\approx 100$ Myr and longer
if the outflows decelerate due to the gravitational potential of the
host halo.  With a typical quasar lifetime of $10-100$ Myr
\citep[e.g.][]{Martini:2004}, the required outflow speed would need to exceed
1000 \kms\ in order for outflows near the peak of an active quasar
phase to explain the observed excess of cool gas at $d\sim 100$ kpc.
With the energetics associated with fast-moving outflows, the gas is
expected to be heated and highly ionized.  If the cool clumps form
within the hot outflows due to efficient cooling 
\citep[e.g.][]{Costa:2015}, then a strong correlation is expected between the
incidence of cool gas and the presence of highly-ionized gas as probed
by either [O\,III] emission \citep[e.g.][]{Greene:2012} or O\,VI/C\,IV
absorption \citep[e.g.][]{Grimes:2009}.  Observations of highly-ionized
gas associated with the low-ionization gas detected in quasar host
halos will provide new insights into the origin of the $\kappa$
vs.\ $L_{\rm bol}$ correlation.

Alternatively, the cool gas could originate in outflows from
star-forming satellites as suggested by a recent simulation study
\citep{Faucher:2016}.  In this scenario, the distance required for
the outflows to travel within the quasar lifetime would be $\sim 10$
kpc, rather than $\sim 100$ kpc, and the required outflow energetics
would be less extreme.  However, the implication would be a more
quiescent satellite environment around low-luminosity quasars in order
to explain a reduced $\kappa$.  Comparing the star formation histories
of satellites around luminous and low-luminosity quasars will provide
a necessary test for this scenario.

\section{Summary and Future Prospects}

Significant progress has been made over the last decade in
characterizing the cool circumgalactic gas in massive halos of
$M_h>10^{12}\,\msol$ at $z\approx 0.2-2$ using absorption
spectroscopy.  This progress is facilitated by the unprecedentedly
large galaxy and quasar samples available in the SDSS spectroscopic
archive.  Both massive galaxies and luminous quasars are rare.  As a
result, finding a background quasar in close projected distances for
absorption-line studies of these rare objects requires a large survey
volume.  The large galaxy and quasar spectroscopic archive helps the
assembly of statistically significant samples of close quasar and
quiescent galaxy pairs and projected quasar pairs.  These pair samples
have enabled systematic studies of low-density gas beyond the nearby
universe.  Key findings from various studies can be summarized as
follows:

\begin{enumerate}

\item Chemically-enriched cool gas of $T\sim 10^4$ K is present in
  massive quiescent halos at $z\sim 0.5$, with a declining annular
  average of covering fraction from $\langle\,\kappa\,\rangle\apg
  15$\% at $d<120$ kpc to $\langle\,\kappa\,\rangle\approx 5$\% out to
  the virial radius $R_{\rm vir}$.  The improved statistics help rule
  out definitively an absence of cool gas in massive quiescent halos
  at $z\sim 0.5$, extending observations of cold gas in local
  early-type galaxies \citep[e.g.][]{Young:2014} to those at
  intermediate redshifts.

\item Strong \MgII\ absorbers of $W_r(2796)>1$ \AA\ produced by
  photo-ionized cool gas are {\it not} uncommon throughout quiescent
  halos from $d<100$ kpc to the virial radius $d\approx R_{\rm vir}$
  and the observed \MgII\ absorbing strength in these halos does not
  depend on either galaxy luminosity or mass.  The lack of correlation
  between $W_r(2796)$ and galaxy properties in quiescent halos
  suggests that the observed cool gas is likely to originate in
  infalling materials from the IGM, rather than outflowing gas from
  these early-type galaxies.

\item The velocity dispersion of \MgII\ absorbing gas around the
  majority ($\approx 90$\%) of massive quiescent galaxies is
  suppressed, at $\approx 60$\% of what is expected from the virial
  motion.  Dissipation is expected if these \MgII\ absorbers originate
  in cool clumps condensed out of the hot halo through thermal
  instabilities and the clumps decelerate due to ram-pressure while
  moving through the hot halo.  In this simple cloud model, the volume
  filling factor of the clumps is small in these massive halos with a
  mean number of $n_{\rm clump}\sim 4$ per sightline in order to
  explain the large scatter found in the $W_r(2796)$ versus $d$
  distribution, and a mean clump mass of $m_{\rm cl}\approx 5\times
  10^4\,\msol$ in order to explain the suppression in velocity
  distribution.

\item While gas metallicity alone is insufficient for distinguishing
  between infalling and outflowing gas due to an unknown degree of
  chemical mixing in the CGM, the observed chemical composition of the
  gas offers important clues for the chemical enrichment history.  The
  chemical composition of cool halo gas at $d\apl 100$ kpc from
  massive quiescent galaxies displays an elevated iron abundance level
  that differs from an $\alpha$-element enhancement typically found in
  star-forming galaxies and in the IGM.  The observed Fe/Mg ratio
  implies a fractional contribution of SNe Ia to the total (Type Ia
  and core-collapse combined) of $f_{\rm Ia}\approx 15-20$\% in these
  inner massive halos.  Beyond $d\approx 100$ kpc, the observed Fe/Mg
  ratio recovers the typical $\alpha$-element enhanced level.

\item There exists a strong correlation between the cool halo gas
  covering fraction $\kappa$ in quasar host halos and quasar
  bolometric luminosity $L_{\rm bol}$, leading to a surge of cool gas
  in halos about luminous quasars at both low and high redshifts.  The
  strong $\kappa$--$L_{\rm bol}$ correlation suggests a physical link
  between cool gas content on scales of 100 kpc and quasar activities
  on sub-parsec scales, but interpreting this strong correlation
  remains challenging.  The primary difficulty lies in the relatively
  short quasar lifetime of $\approx 10-100$ Myr in comparison to the
  long dynamical time necessary to move gaseous clouds over a large
  distance of $\approx 200$ kpc.  Direct imaging of quasar-driven
  outflows and observations of highly-ionized gas associated with cool
  gas at $d\approx 200$ kpc are necessary to establish direct
  connections between outflows and cold gas detected at large
  distances.

\end{enumerate}

Together, these findings suggest that infalling clouds from external
sources are likely a dominant source of cool gas detected at $d\apg
100$ kpc from massive quiescent galaxies.  The origin of the gas
  closer in is currently less certain, but SNe~Ia driven winds appear
  to contribute significantly to cool gas found at $d< 100$ kpc.  In
contrast, cool gas observed at $d\apl 200$ kpc from luminous quasars
appears to be intimately connected to the on-going quasar activities.
The observed strong correlation between cool gas covering fraction in
quasar host halos and quasar bolometric luminosity remains a puzzle.

With new instruments and new survey data becoming available,
continuing progress is expected in a number of areas over the next few
years for a better understanding of the CGM in massive halos.  In
particular, spatially-resolved observations of quasar outflows in the
inner $10-30$ kpc region, combined with absorption-line kinematics at
$\sim 100$ kpc from the quasar, will provide key insights into the
strong correlation between $\kappa$ and $L_{\rm bol}$ in quasar host
halos.  Integral field unit (IFU) spectrographs available on large
ground-based telescopes provide a powerful tool for imaging quasar
outflows based on observations of high-ionization lines.

In addition, while measurements of chemical compositions provide a
unique constraint for the physical origin of chemically-enriched gas
in massive quiescent halos, measurements of $N({\rm HI})$ are
necessary for direct comparisons between observations and
state-of-the-art cosmological simulations.  The Cosmic Origins
Spectrograph \citep[COS;][]{Green:2012} on board the {\it Hubble Space
  Telescope} provides the spectral coverage necessary for $N({\rm
  HI})$ measurements at $z\apl 1$.  With an increasing number of
$z\approx 0.5$ LRGs found near the sightline of a UV bright QSO, there
will soon be a statistical sample of massive quiescent halos with
known $N({\rm HI})$ at different projected distances for testing
simulation predictions.

Furthermore, understanding the roles of satellites and satellite
interactions in producing chemically-enriched cool clumps in
low-density halos requires deep galaxy survey data in quasar fields.
With several wide-field integral field spectrographs being installed
on ground-based telescopes, deep galaxy survey data in a large number
of quasar fields will soon be available for systematic studies of the
galaxy environments of kinematically complex absorbers.

Finally, little is known regarding the CGM properties and galactic
environments of massive starburst galaxies with $M_{\rm star}\apg
10^{11}\,\msol$ \citep[e.g.][]{Borthakur:2013}.  Although these
galaxies are very rare, contributing to roughly 10\% of the massive
galaxy population (with the rest being quiescent LRGs), they are
intrinsically UV luminous and massive stars in these galaxies serve as
the backlight for probing the internal star-forming ISM in front of
the massive young stars.  With numerous large-scale galaxy surveys
expected in the coming years \citep[e.g.][]{Zhu:2015}, combining
intrinsic absorption-line observations with absorption spectroscopy
along transverse sightlines \citep[e.g.][]{Rubin:2010G, Kacprzak:2014}
will be feasible for a statistically significant sample of massive
starburst galaxies.  These new data will offer an important empirical
understanding of the impact of starbursts on the CGM in massive halos.

\begin{acknowledgement}
The author wishes to thank Sean Johnson, Rebecca Pierce, Michael
Rauch, and Fakhri Zahedy for providing helpful input and comments.  In
preparing this review, the author has made use of NASA's Astrophysics
Data System Bibliographic Services.
\end{acknowledgement}

\end{document}